\documentclass[aps,prb,twocolumn,groupedaddress,longbibliography,
floatfix,superscriptaddress]{revtex4-2}

\pdfoutput=1
\usepackage[english]{babel}
\usepackage{graphicx,graphics,epsfig,subfigure,times,bm,bbm,amssymb,amsmath,amsthm,mathrsfs,MnSymbol}
\usepackage{gensymb}
\usepackage{amsfonts}
\usepackage{float}
\usepackage{ragged2e}
\usepackage[matrix,frame,arrow]{xypic}
\usepackage[pdfstartview=FitH]{hyperref}

\hypersetup{
colorlinks=true,       
linkcolor=red,          
citecolor=blue,        
filecolor=magenta,      
urlcolor=blue,           
runcolor=cyan
}
\usepackage[pdftex]{color}
\usepackage{braket}  
\usepackage{enumerate}
\usepackage[normalem]{ulem}
\usepackage[usenames,dvipsnames]{xcolor}
\usepackage{multirow}
\usepackage{mathtools}
\usepackage{bbm}
\usepackage{titletoc}

\definecolor{orange}{rgb}{1,0.5,0}

\newcommand{\RNum}[1]{\uppercase\expandafter{\romannumeral #1\relax}}

\newcommand{\ignore}[1]{}
\usepackage{geometry}

\ignore{
\documentclass[eprintnumbers,amsmath,amssymb,onecolumn,a4paper,caption ]{article}
\usepackage{amsfonts}
\usepackage{amssymb}
\usepackage{mathrsfs}
\usepackage{mathbbold}
\usepackage{bbm}
\usepackage{mathrsfs}
\usepackage{dcolumn}
\usepackage{bm}
\usepackage{times,epsfig,amssymb,amsmath}
\usepackage{float}
\usepackage{subfigure}
\usepackage{geometry}\geometry{left=2.5cm,right=2.5cm,top=3cm,bottom=3cm}

\usepackage{color}

}

\begin{document}
\title{{Fast scrambling dynamics and many-body localization transition in an all-to-all disordered quantum spin model}}

\author{Shang-Shu~Li}
\affiliation{Beijing National Laboratory for Condensed Matter Physics, Institute of Physics, Chinese Academy of Sciences, Beijing 100190, China}
\affiliation{School of Physical Sciences, University of Chinese Academy of Sciences, Beijing 100190, China}
\author{Rui-Zhen~Huang}
\email{huangrzh@icloud.com}
\affiliation{Kavli Institute for Theoretical Sciences, University of Chinese Academy of Sciences, Beijing 100190, China}

\author{Heng~Fan}
\email{hfan@iphy.ac.cn}
\affiliation{Beijing National Laboratory for Condensed Matter Physics, Institute of Physics, Chinese Academy of Sciences, Beijing 100190, China}
\affiliation{School of Physical Sciences, University of Chinese Academy of Sciences, Beijing 100190, China}
\affiliation{CAS Center for Excellence in Topological Quantum Computation, UCAS, Beijing 100190, China}
\affiliation{Songshan Lake Materials Laboratory, Dongguan, Guangdong 523808, China}

\begin{abstract}
We study the quantum thermalization and information scrambling dynamics of an experimentally realizable quantum spin model with homogeneous XX-type all-to-all interactions and random local potentials. We identify the thermalization-localization transition by changing the disorder strength, under a proper all-to-all interaction strength. The scrambling dynamics in the localization phase shows novel behaviors distinct from that of local models. The operator scrambling grows almost equally fast in both phases. In the thermal phase, we show there exhibits fast scrambling without appealing to the semi-classical limit.  
We also briefly discuss the experimental realization of the model using superconducting qubit quantum simulators.  
\end{abstract}
\maketitle{\tiny}

\section{Introduction}Isolated out-of-equilibrium quantum many-body systems tend to become thermal and serve as their own thermal bath due to the interaction, known as quantum thermalization~\cite{deutsch_quantum_1991, srednicki_chaos_1994}. Recently, quantum chaos and information scrambling~\cite{hosur_chaos_2016,sekino_fast_2008, shenker_black_2014, maldacena_bound_2016,nahum_operator_2018,yan_information_2020,rakovszky_diffusive_2018,couch_speed_2020,xu_locality_2019,mcginley_slow_2019} in thermalized systems have attracted great interest for their importance in understanding the non-equilibrium dynamics in the strongly interacting system and quantum gravity.  Information scrambling describes how local information spreads to other degrees of freedom in the quantum chaotic system under the unitary evolution. It is fundamental for studying the dynamics of black holes and quantum information processing.

A particular recent focus is the speed limit of information scrambling dubbed fast scrambling conjecture~\cite{maldacena_bound_2016,lashkari_towards_2013}, in which the scrambling time $t_s$ for information spreading to the entire system satisfies
\begin{equation}\label{scr_t}
t_s\sim \log (N),
\end{equation}
where $N$ is the system size. Black holes are known as the fast scrambler in nature that saturates the upper bound. Another celebrated model that exhibits fast scrambling is the Sachdev-Ye-Kitaev (SYK) model~\cite{sachdev, Kitaev, maldacena_remarks_2016, kobrin_many-body_2021}, which has been proved to be holographically dual to quantum gravity. On the other hand, the rapid development of highly controlled quantum simulators enables us experimentally study information scrambling~\cite{li_measuring_2017,garttner_measuring_2017,swingle_measuring_2016,blocher_measuring_2020,lewis-swan_unifying_2019,joshi_quantum_2020,vermersch_probing_2019,braumuller_probing_2021}. Thus, it is interesting to find other experimental realizable quantum models that exhibit fast scrambling.

Generally, accessing fast scrambling in quantum many-body systems needs non-local interaction and chaotic dynamics~\cite{marino_cavity-qed_2019, bentsen_treelike_2019,belyansky_minimal_2020, li_fast_2020,bentsen_fast_2019}. Quantum systems with short-range or even power-law long-range interactions are prevented from fast scrambling for the existence of light-cone~\cite{lieb_finite_1972, hastings_spectral_2006, cheneau_light-cone-like_2012, richerme_non-local_2014, foss-feig_nearly_2015, kuwahara_absence_2021, else_improved_2020,chen_finite_2019}.  Recently, several models are proposed by using quasi-random all-to-all interaction~\cite{marino_cavity-qed_2019}, tree-like interaction~\cite{bentsen_treelike_2019} or the combination of all-to-all and local interactions~\cite{belyansky_minimal_2020, li_fast_2020} to reach experimentally accessible fast scrambling. { However, most of these studies manifest fast scrambling only when considering the semi-classical limit and the direct signature Eq.~(\ref{scr_t}) is absent for quantum spin models. Studies that directly demonstrate a fully quantum model exhibits fast scrambling without appealing to semi-classical limits are still lacking, except the SYK model~\cite{kobrin_many-body_2021}}.

In this paper, we study an all-to-all quantum spin model with local quenched disorders. A weak disorder strength can induce a thermal phase in the quantum many-body system, simultaneously the all-to-all interaction prevents the existence of light-cone that limits the information spreading. {Using the state-of-the-art tensor network method, we demonstrate this model exhibit fast scrambling without taking the semi-classical limit from the scaling of the scrambling time with respect to the system size $N$.}

We first study the thermalization-localization transition between the thermal and many-body localization (MBL) phases ~\cite{pal_many-body_2010, oganesyan_localization_2007, huse_phenomenology_2014,nandkishore_many-body_2015, chandran_constructing_2015, serbyn_criterion_2015, bardarson_unbounded_2012, vosk_theory_2015, serbyn_criterion_2015, morningstar_many-body_2020, luitz_many-body_2015, potter_universal_2015, dumitrescu_scaling_2017, modak_many-body_2015} to locate the critical point. Then, we study the scrambling dynamics by using the out-of-time-ordered correlation (OTOC)~\cite{larkin_otoc_1969, Kitaev} . We show that the scrambling dynamics is not slowed down even in the MBL phase, however it can be distinguished by later time behavior to that in the thermal phase. We identify the fast scrambling dynamics without taking the semiclassical limit in the thermal phase.  We also present the scrambling dynamics for different interaction strengths and discuss the relationship to the time scale for the scrambling dynamics at infinite temperatures.

The rest of this paper is organized as follows. In Sec.~\ref{model}, we give of brief introduction of the model investigated. Sec.~\ref{transition} gives the phase transition between the thermal and MBL phase and Sec.~\ref{Scrambling} studies the general properties of scrambling dynamics in these two phases. In Sec.~\ref{fs} we study the fast scrambling dynamics at infinite temperature. Sec.~\ref{summary} gives a summary of our conclusions. In Appendix, the convergence of our numerical method and the time scale for infinite temperature ensemble are discussed.
\begin{figure}[t]
  \centering

  \includegraphics[width=0.46\textwidth=]{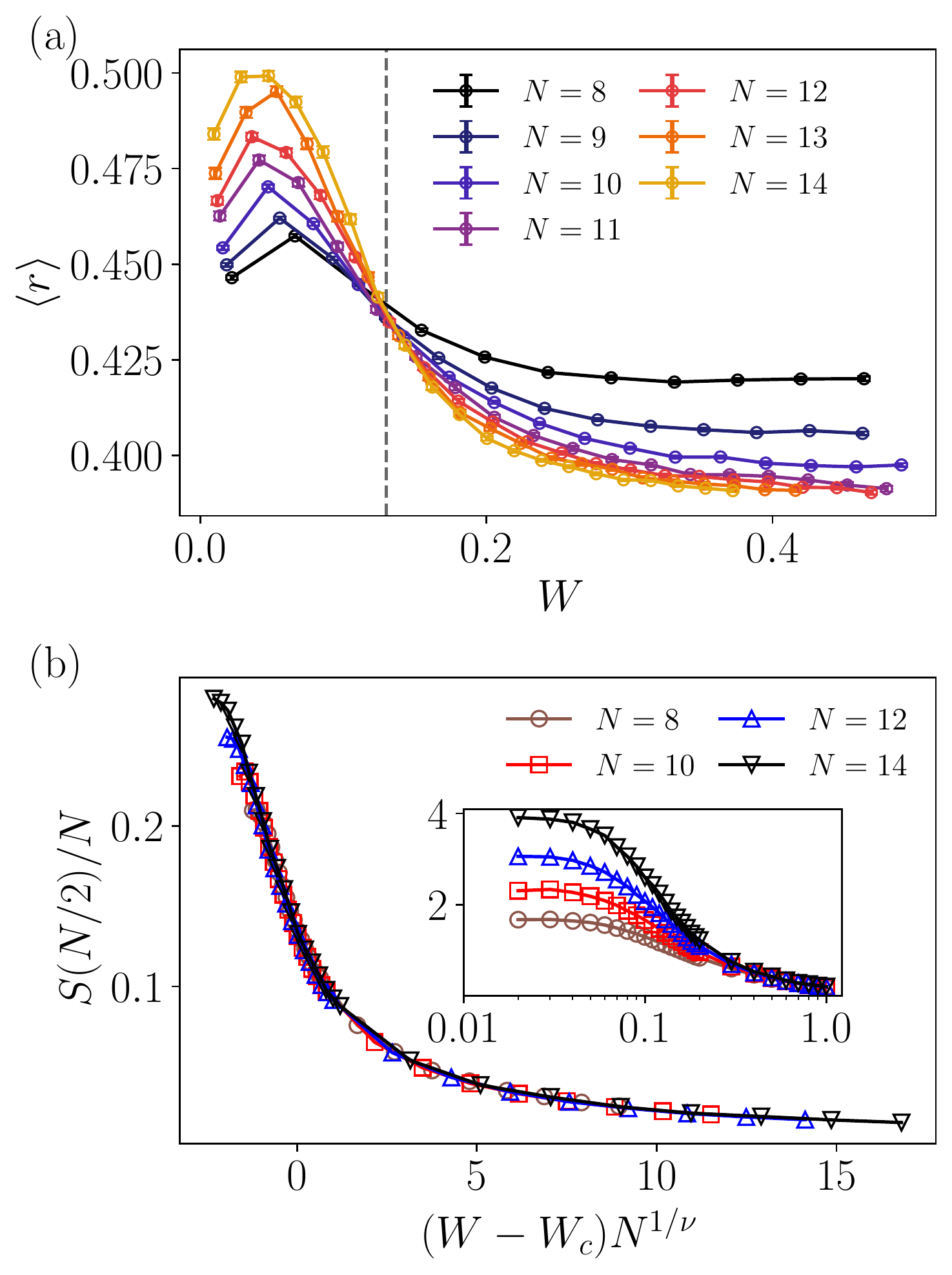}
  \caption{Thermalization-localization transition. (a) The adjacent gap ratio $\langle r \rangle$ as a function of the disorder strength $W$. The dashed line denotes the thermalization-localization transition point. (b) The entanglement entropy v.s. the disorder strength before (inset) and after rescaling. The disorder average is taken over $10000$ samples for $N=8$ and $500$ for $N=14$. The middle $1/3$ spectra are taken for $\langle r \rangle$ and central 20 states for $S(N/2)$. The error bars are smaller than the points.}
  \label{fig:MBL}
  \end{figure}

  \section{Model}\label{model}
We consider a quantum spin lattice model with fully-connected XX-type interactions and random on-site transverse magnetic fields,
\begin{equation}\label{Hamil}
\hat{H} = \frac{J}{2N^{\alpha}}\sum_{i<j} \left( \hat{\sigma}^{x}_i \hat{\sigma}^{x}_j + \hat{\sigma}^{y}_i \hat{\sigma}^{y}_j \right) + \sum_i h_i\hat{\sigma}^x_i
\end{equation}
\begin{figure*}[t]
  \centering
  \includegraphics[width=0.9\textwidth=]{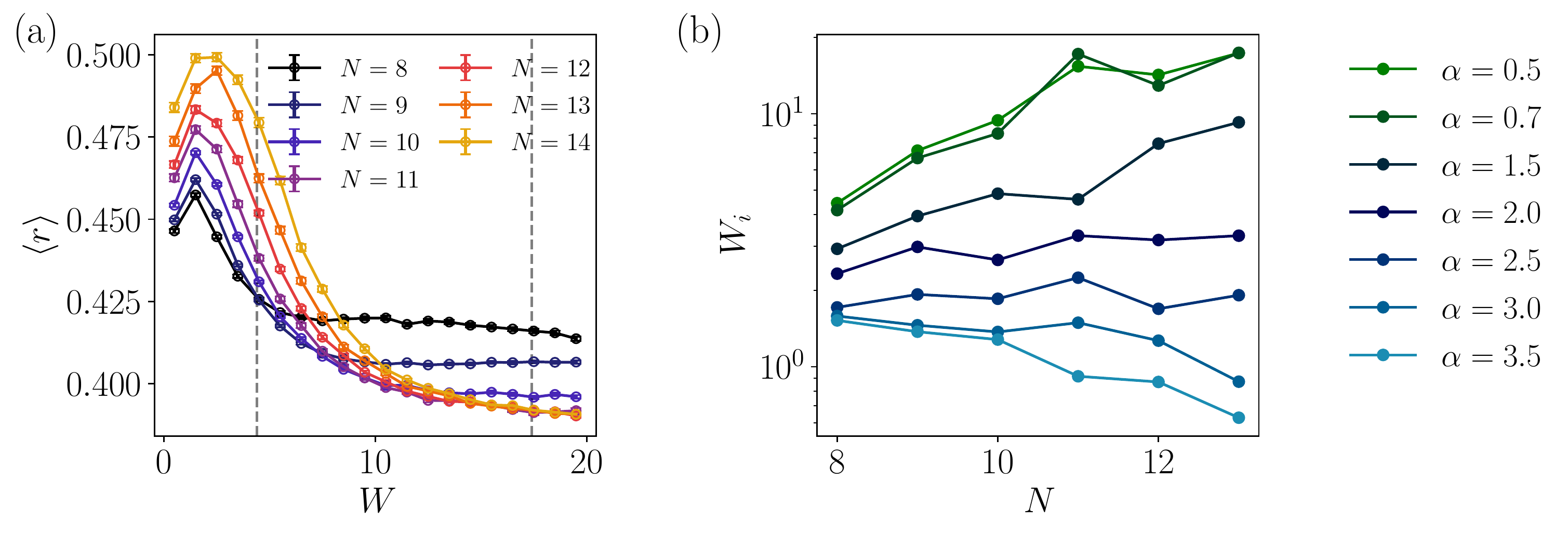}
  \caption{(a) Adjacent ratio gap as a function of disorder strength, under the scaling $\alpha=0.5$. The dashed lines mark out the region where the intersecting points shift with system size. (b) The intersecting points $W_i$ with different ${\alpha}$. These $W_i$ converge to a finite value as increasing $N$ under ${\alpha}\sim2$.}
  \label{fig:extfig1}
  \end{figure*}
where $\hat{\sigma}_i^{x,y}$ denotes the Pauli matrix, $J$ is the coupling constant and $h_i$ serves as uniformly distributed random fields $ h_i \in [-W,W] $. Without disorder, i.e., $W=0$, $\hat{H}$ reduces to the exactly solvable Lipkin-Meshkov-Glick model~\cite{lipkin_validity_1965, ribeiro_thermodynamical_2007}. The disordered field is introduced to render the model non-integrable.  The coupling strength $J$ is normalized by a factor $1/N^{\alpha}$ with respect to the total number of sites $N$ and is fixed as $J=1$ in this paper. Thus, it is clear that $\alpha$ governs the disorder strength relative to the all-to-all interaction and controls the localization phase transition. On the other hand, the value of $\alpha$ is shown crucial to realizing fast scrambling Eq.~(\ref{scr_t}) in this type of all-to-all model. For example, Ref.~\cite{belyansky_minimal_2020, li_fast_2020} suggest use $\alpha = 0.5$ and Ref.~\cite{yin_bound_2020} proves that for $\alpha > 0.5$ there is no fast scrambling. 
In addition, the model Eq.~(\ref{Hamil}) hosts a global $Z_2$ symmetry, $\hat{\sigma}^y   \rightarrow - \hat{\sigma}^y$. In the numerical study of thermalization-localization phase transition, the calculation is done in the $Z_2$ parity even sector.

Our model is directly achievable in state-of-the-art quantum simulation platforms, such as superconducting qubit quantum simulators~\cite{song_10-qubit_2017,song_generation_2019,xu_probing_2020}. With a cluster of qubits coupled to a single cavity resonator, the all-to-all XX-type interaction can be conveniently generated and accurately controlled. The on-site random transverse fields are also realizable by applying microwaves driving on each qubit with adjustable amplitudes and frequencies. The scrambling dynamics may be probed by statistical correlation of random measurements~\cite{joshi_quantum_2020,vermersch_probing_2019} or considering the fidelity OTOC~\cite{lewis-swan_unifying_2019}.

\section{Thermalization-localization transition}\label{transition} 
We first study the chaotic properties of the model, Eq.~(\ref{Hamil}), under the disordered field $h_i$. With the increase of the disorder strength, the quantum dynamics becomes slow and can evolve to a localized phase, i.e., the ergodicity is breaking. Generally, the thermal-localization transition can be characterized by the statics of energy levels $E_n$. In the thermal phase, it shows Wigner-Dyson distribution, while becomes Poisson in the localized phase. Numerically, the mostly used quantity characterizing the two phases is the ratio of adjacent gap~\cite{atas_distribution_2013}, $\bar{r}_n=\min(r_n, 1/r_n)$, where $r_n=(E_{n+1}-E_n)/(E_n-E_{n-1})$. The average value $r=\langle \bar{r} \rangle$ with respect to energy levels $n$ satisfies $r=0.38$ corresponding to the Poisson distribution, while $r = 0.53$ corresponding to the Wigner-Dyson distribution. The transition can also be reflected from the half-system bi-partite entanglement entropy $S(N/2)=-\text{Tr}(\rho_{N/2}\ln \rho_{N/2})$ for the states at the middle of the energy level. It dramatically changes from the volume law in the thermal phase to the area law in the localized phase.

We first use exact-diagonalization (ED) to study the localization transition via the finite-size scaling of the disorder averaged adjacent gap ratio $\langle r \rangle$. It has been argued that the localization transition in all-to-all models can only happen with a proper rescaling of the disorder strength ${\alpha}=2$~\cite{modak_many-body_2020,tikhonov_many-body_2018,gornyi_spectral_2017,gopalakrishnan_instability_2019}. We confirm this argument from the numerical results. For ${\alpha}=2$, the system undergoes a thermal-localization transition, as shown in Fig.~\ref{fig:MBL} (a). We find that different $r$ curves cross at the transition point $W_{c} = 0.14(2)$ for different system sizes. Away from the transition point, it approaches the theoretical value 0.38 or 0.53 in the localized and thermal phases respectively. For the case ${\alpha} \neq 2$,
the transition point $W_c$ tends to be smaller as increasing the system size for the ${\alpha}>2$ case, while becomes larger when ${\alpha}<2$. Fig. \ref{fig:extfig1} (a) presents the level statics with ${\alpha}=0.5$. We observe that the intersecting points $W_i$ of consecutive curves quickly shift when increasing the system size. This may indicate the absence of the MBL phase in the thermodynamic limit. Furthermore, by studying the shift of intersecting points under different ${\alpha}$, we find that the intersecting points depend on system size strongly except ${\alpha}\sim2.0$. Specifically, in Fig. \ref{fig:extfig1} (b), the intersecting points for $r$ flow to larger or smaller values when increasing the system size for ${\alpha}<2.0$ or ${\alpha}>2.0$ respectively. Therefore a localized phase may become thermal for larger systems, or vice versa. A systematic study of whether it will flow to infinite (zero) for ${\alpha}<2.0$ (${\alpha}>2.0$) is beyond the study of our work, which may be only accessible to renormalization group based approaches.

Furthermore, we perform a finite-scaling analysis of $S(N/2)$ in Fig.~\ref{fig:MBL}(b). The entanglement entropy satisfies volume and area law in the thermal and localized phases respectively, as expected (see insect of Fig.~\ref{fig:MBL}(b)). We then perform a data collapse according to the scaling form $S(N/2)=N f((W-W_c)N^{1/\nu})$, where $\nu$ is the critical exponent. The curves of different sizes accurately collapse onto each other with $W_c=0.14(2)$ and $\nu=0.89(2)$. The critical point obtained in this way is consistent with that determined by the crossing of $\langle r \rangle$, demonstrating that the critical properties obtained here are accurate and reliable. These results provide the first numerical evidence supporting the existence of thermal-localization transition in all-to-all models.

\section{Scrambling dynamics in the thermal and MBL phases}\label{Scrambling}
After identifying the thermal and localized phases, we continue to study the information scrambling in both phases. The most important quantity is the squared commutator of two local operators $\hat{W}_i$ and $\hat{V}_j$ at infinite temperature
\begin{equation}
C(r,t)=\frac{1}{2}\langle[\hat{W}_0(t), \hat{V}_r]^{\dagger}[\hat{W}_0(t), \hat{V}_r]\rangle,
\end{equation}
where $\hat{O}_r(t)=e^{i\hat{H}t}\hat{O}_r e^{-i\hat{H}t}$ denotes a Heisenberg operator. The commutator $C(r,t)$ is closely related to the OTOC $F(r,t) = \langle \hat{W}_0^\dagger(t) \hat{V}_r(0)^{\dagger} \hat{W}_0(t) \hat{V}_r(0) \rangle$ according to $C(r,t) = \left(1-\mathrm{Re}[F(r,t)])\right)$. It provides a clear insight to the scrambling dynamics. That is, a local operator under time evolution grows non-local with time and becomes non-commuting with operators at other sites. It is expected that $C(r, t)$ grows exponentially before saturation for fast scramblers~\cite{maldacena_bound_2016}, analogue to the classical butterfly effect. In the following we focus on $C(r,t)$ of $\hat{\sigma}^z$, namely $\hat{W}=\hat{V}=\hat{\sigma}^z$.

\begin{figure}[t]
  \centering
  \includegraphics[width=0.5\textwidth=]{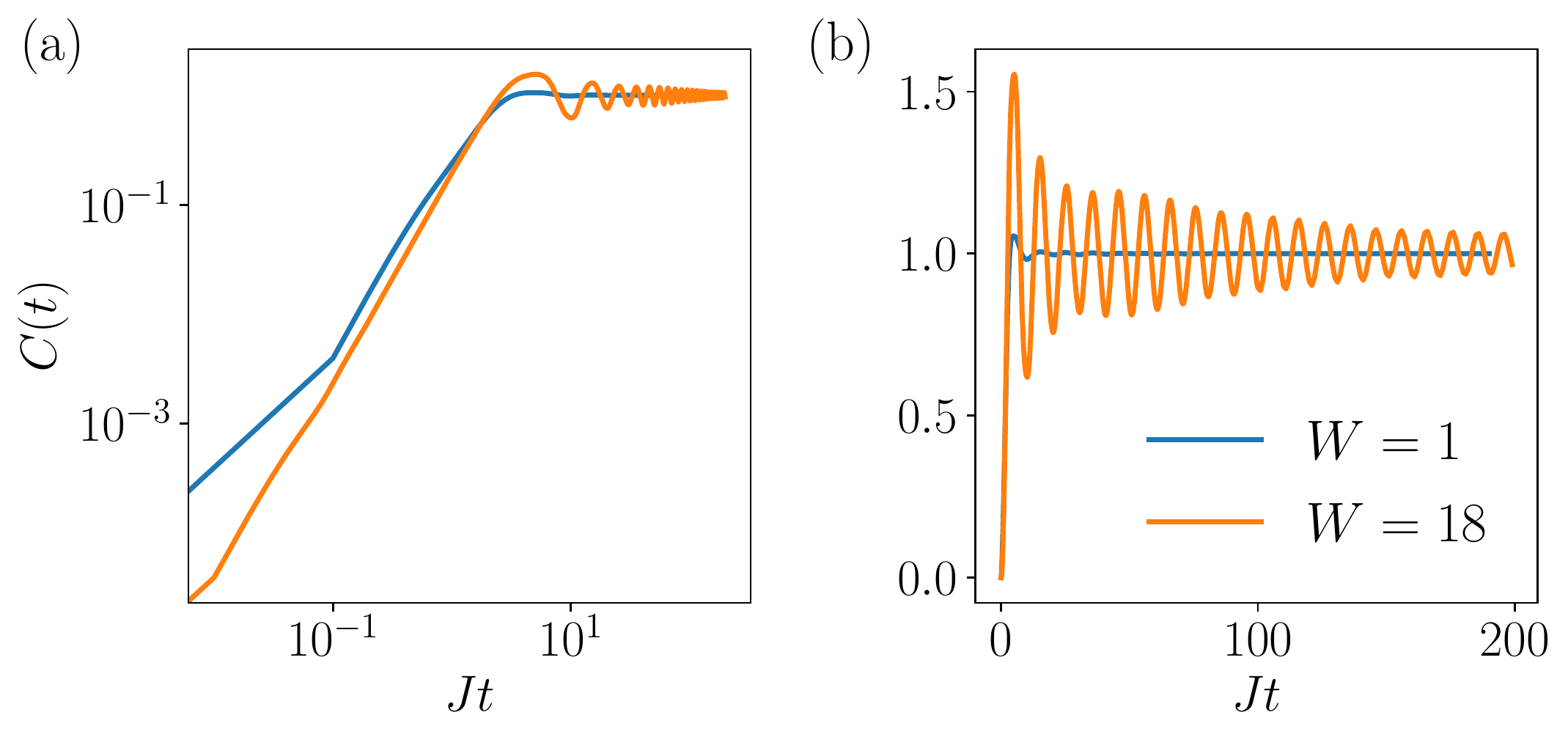}
  \caption{ OTOC in the thermal ($W=1$) and MBL ($W=18$) phases for a finite system with $N=10$, and $\alpha=0.5$. (a) Early time behavior of $C(t)$ in a log-log plot, which shows a power-law like growth. The disorder average is done over 50 samples. (b) Long time behavior of $C(t)$ in the thermal and MBL phase.}
  \label{fig:fig3}
  \end{figure}
 
  \begin{figure*}[t]
    \centering
    \includegraphics[width=0.9\textwidth=]{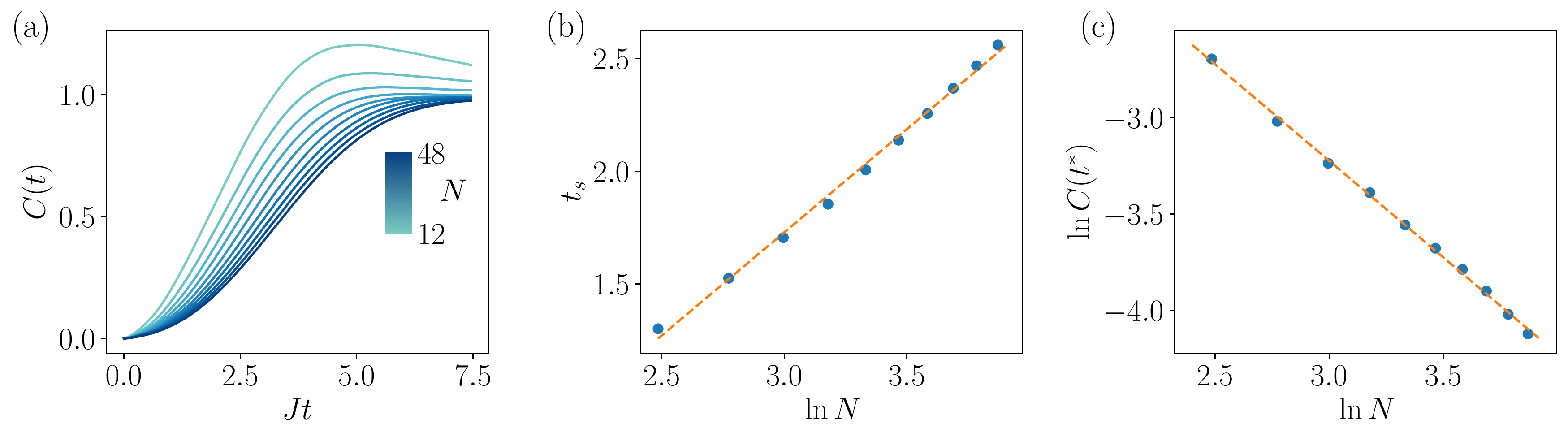}
    \caption{Scrambling dynamics for different system sizes. (a) Time evolution of $C(t)$ for various system sizes $N=12,16,20,...48$ with $\alpha=0.5$. The disorder strength is $W=2$ and the system is in the thermal phase. (b) The scrambling time $t_s$ versus the system size $N$. The dashed line is the linear fit for $\alpha=0.5$, indicating a logarithmic growth of $t_s$. (c) The power-law decay of the initially built correlation $C(t^*)$ versus the system size $N$.  The data is taken from (a) with $t^*=0.5$.  }
    \label{fig:fig4}
    \end{figure*}
Since there is no spatial difference in our all-to-all model, we consider the mean commutator $C(t)$ for all $r>0$, where $C(t)= \frac{1}{N-1}\sum_{r>0} C(r,t)$.  The $C(t)$ of our model presents unique time evolution behaviors as shown in Fig.~\ref{fig:fig3} (a-b).   It can be seen that even in the localized phase the early-time growth of the commutator is not arrested in our model. The mean commutator grows almost equally fast to saturation for both phases in an approximate power-law form.   The difference between the two phases lies in the late time behavior, where $C(t)$ is almost invariant in the thermal phase while periodically oscillates for long times in the MBL phase, see Fig.~\ref{fig:fig3}(b). The late-time oscillation of $C(t)$ indicates the breakdown of the chaotic dynamics in the MBL phase. 
The early and late time behaviors of $C(t)$ in the MBL phase of our model are different from that in the local MBL phase~\cite{xu_accessing_2020,fan_out--time-order_2017, huang_out--time-ordered_2017, sahu_scrambling_2019,lee_typical_2019,he_characterizing_2017,chen_out--time-order_2017} in which $C(r,t)$ exhibits a logarithmic light cone and grows in a power-law way for a long time. These novel behaviors result from the interplay between the all-to-all interaction and localization. We conjecture that there may exist non-local integrals of motion beyond the phenomenology model for the conventional MBL phase\cite{chen_out--time-order_2017}. The detailed study of the non-equilibrium dynamics in the all-to-all MBL phase using the effective theory is left for further study.

\section{Fast scrambling}\label{fs}
Now we focus on the fast scrambling dynamics in the thermal phase. We have shown that for a small system the early growth of $C(t)$ is in an approximate power-law form rather than exponential. However, exponential growth is generally expected for the semi-classical system with a large system size $N$ in the large spin-$S$ limit. While it is not guaranteed for quantum models with limited system size and small local Hilbert space~\cite{hashimoto_out--time-order_2017,khemani_velocity-dependent_2018,cotler_out--time-order_2018}.  Nevertheless, we will show that there exists fast scrambling in our model.

To study the fast scrambling dynamics, one needs to resort to larger  system sizes, which are generally inaccessible to the ED study. We utilize the recently developed tensor-network method~\cite{xu_accessing_2020, bentsen_treelike_2019,zhou_operator_2020} based on the matrix product operator representation of Heisenberg operators and the time-dependent variational principle~\cite{haegeman_time-dependent_2011, haegeman_unifying_2016} (TDVP-MPO) to study the OTOC in the early time growth region. {The accuracy of the TDVP-MPO method is 
determined by the entanglement between operators rather than the entanglement in pure states. It was demonstrated the operator entanglement is quite limited at early times and the TDVP-MPO approach can accurately capture the operator entanglement properties~\cite{zhou_operator_2020, leviatan_quantum_2017,xu_accessing_2020}. This enables us to study larger system, $O(50)$ spins, in this work. The convergence of our numerical results is discussed in Appendix.~\ref{append_converge}}

We first present the $C(t)$ with various system sizes under $\alpha=0.5$ in Fig.~\ref{fig:fig4} (a). One can see the curves shift clearly when increasing the system size, which is similar to that in the SYK model~\cite{kobrin_many-body_2021}. The most direct and important criterion for the fast scrambling comes from $t_s \sim \mathrm{log} \left(N\right)$, where $t_s$ is defined as the timescale at which $C(t_s)$ reaches a saturated value ($C(t_s) \simeq 0.3$ in our study~\cite{note}).We show that our numerical results in Fig.~\ref{fig:fig4}(b) are consistent with this relation. It provides strong numerical evidence for the fast scrambling dynamics in our model. Note that this direct numerical evidence for Eq.~(\ref{scr_t}) is usually absent in previous studies due to the limited system size, except in the SYK model~\cite{kobrin_many-body_2021}. We also verify another necessary condition for fast scrambling, namely the initially built correlation is at most algebraically small ($C\sim N^{-\gamma}$) rather than exponential small in $N$~\cite{bentsen_treelike_2019, belyansky_minimal_2020}. In Fig.~\ref{fig:fig4} (c), we plot the $C(t)$ at a fixed early time $t^*=0.5$ for different system sizes. In the log-log scale, the linear fit demonstrates a power-law decay of $C(t^*)$ versus $N$. For the case of $\alpha \neq 0.5$, we find the fast scrambling is related to the time scale and is ambiguous at infinite temperature. We leave the discussion for this case in the Appendix~\ref{other_alpha}.

Additionally, we give a semi-classical calculation for our model with a much larger system size.
\begin{figure*}[t!]
  \centering
  \includegraphics[width=0.62\textwidth=]{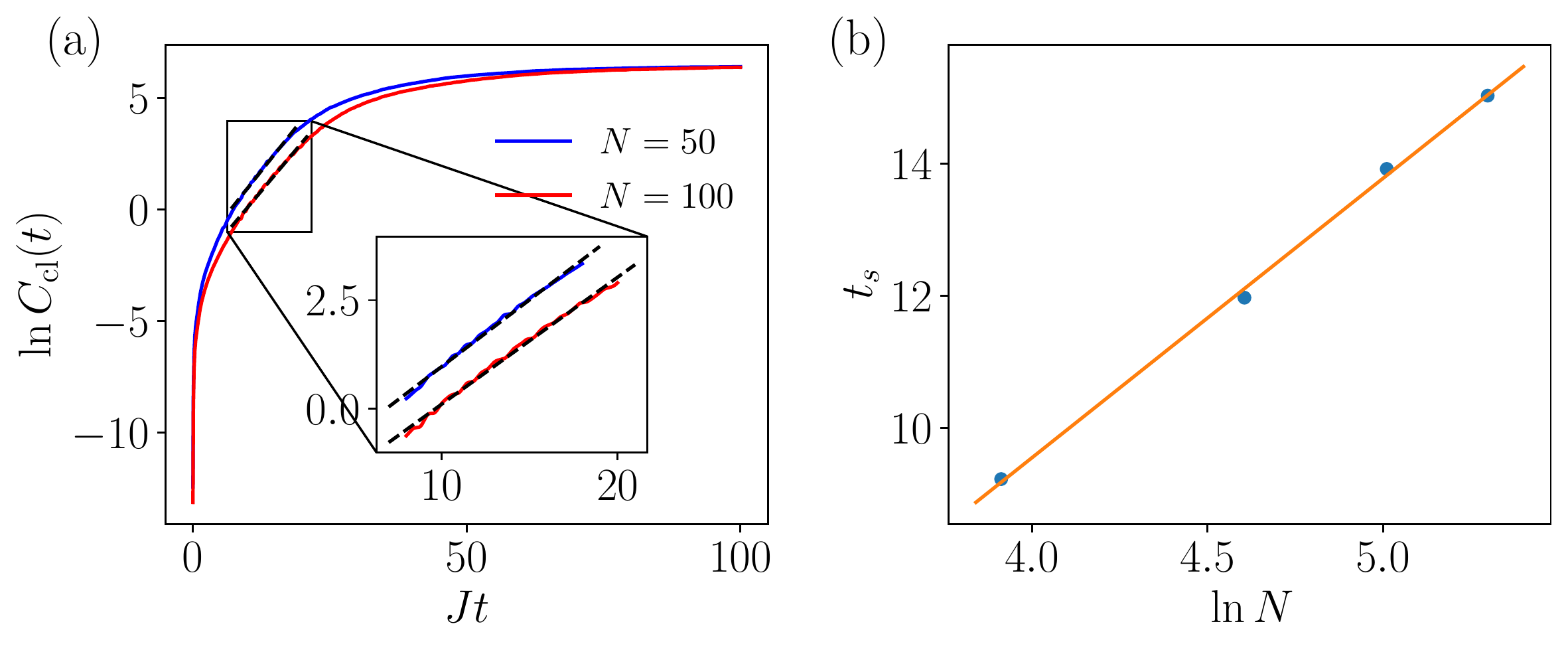}
  \caption{Semiclassical numerics for fast scrambling dynamics with $\alpha=0.5$ and $W=1$. (a) The exponential growth of averaged sensitivity $C_{\text{cl}}(t)$ at intermediate-time region. (b) Fast scrambling in the semi-classical limit, with system size up to $N=200$.
  }
  \label{fig:extfig4}
  \end{figure*}
That is, when the spin length $S$ tends to infinite, the spin operator $\hat{\textbf{S}}$ can be replaced by the classical angular momentum vector $\textbf{S}$. Correspondingly, the commutator in the Heisenberg equation is replaced by the Poisson bracket. For large S models, the results of thermal-localization transition derived for the spin-1/2 case are no longer applicable. We thus choose a value of $W$ that renders the semi-classical dynamics chaotic and easy to numerically integrate.  Following Ref.~\cite{marino_cavity-qed_2019, cotler_out--time-order_2018}, we calculate the semi-classical averaged sensitivity
\begin{equation}\label{semi_C}
C_{\text{cl}}(r, t)=\frac{1}{S^2}\left\langle\left(\frac{dS^z_r(t)}{d\phi}\right)^2 \right\rangle,
\end{equation}
where $\phi$ is an initial small rotation of spin 0 about the $z$ axis.  $C_{\text{cl}}(r, t)$ can be seen as the semi-classical version of  $C(r,t)$. The average in Eq.~(\ref{semi_C}) is done for an initial ensemble that each spin randomly lies in the $x-y$ plane, for each disorder realization. The growth of the disorder averaged $C_{\text{cl}}(t)$ is shown in Fig.~\ref{fig:extfig4} (a), where an additional average is done over sites $r>0$. We can clearly see three stages of the growth. Following the first stage of rapid power-law growth, there exists an exponential growth of $C_{\text{cl}}(t)$ corresponding to the Lyapunov region. At last stage the growth of averaged sensitivity slows down and tends to saturate. We extract the scrambling time with the saturated value $C_{\text{cl}}(t)=1$ in the second stage, up to the system size $N=200$. It is shown in  Fig.~\ref{fig:extfig4} (b) that the semi-classical dynamics also presents a fast scrambling.

\section{Summary and outlook}\label{summary}
In summary, we have studied the thermalization-localization transition and scrambling dynamics in a disordered all-to-all model. We characterize the transition under a proper scaling of the interaction strength $\alpha$. We study the scrambling dynamics in both thermal and MBL phase. In the MBL phase, the scrambling is almost the same fast as the thermal phase while exists long time oscillation at late time. We confirm the existence of fast scrambling in our model without appealing to the semi-classical limit. 

There remain several open problems that need further study. A more detailed study of the novel scrambling dynamics in the MBL phase of all-to-all or other types of globally interacting models is needed. Moreover, with both global and local interactions, there may exist a transition between local and global MBL phases by tuning the strength of the global and local interaction. Further works may also include the study of the scrambling dynamics at a finite temperature. The lack of exponential growth in finite-size spin-$1/2$ systems may be further addressed by considering the models without conservation laws such as Floquet systems~\cite{chen_out--time-order_2017} with all-to-all interactions.

\begin{acknowledgments} 
    The tensor network simulation is performed using the ITensor library~\cite{itensor}. We thank Cheng Peng for helpful discussions about fast scrambling dynamics. This work is supported by the National Natural Science Foundation of China (Grants Nos. 12047554, 11934018), China Postdoctoral Science Foundation (Grant No. 2020T130643), the Fundamental Research Funds for the Central Universities, Strategic Priority Research Program of Chinese Academy of Sciences (Grant No. XDB28000000), Scientific Instrument Developing Project of Chinese Academy of Sciences (Grant No. YJKYYQ20200041) and Beijing Natural Science Foundation (Grant No. Z200009).
\end{acknowledgments}

\appendix
\section{Convergence of numerical results}\label{append_converge}
In this section, we briefly discuss the tensor network simulation method used in our work and the convergence of our numerical results. We utilize the recently proposed TDVP-MPO method~\cite{bentsen_treelike_2019, zhou_operator_2020} to simulate the scrambling dynamics. The commutator $C(r, t)$ is calculated by the following relation
\begin{equation}
C(r, t)=1-\mathrm{Re} F(r, t)
\end{equation}, where $F(r, t)$ is defined as
\begin{equation}\label{F}
F(r, t) = \langle \hat{\sigma}^z_0(t) \hat{\sigma}^z_r \hat{\sigma}^z_0(t) \hat{\sigma}^z_r \rangle
\end{equation}
for $ \hat{\sigma}^z_i$ operator. The TDVP-MPO method for calculating Eq.~(\ref{F}) is by directly evolving the operator using the Heisenberg equation. The time evolved operators are represented as MPO with a finite bond dimension. Then one can map the MPO to matrix product state (MPS) by vectorizing each local operator and employing the standard TDVP algorithm \cite{haegeman_time-dependent_2011, haegeman_unifying_2016} to evolve the MPS in time. The effective Hamiltonian for this MPS is constructed as the super-operator $H\otimes I-I\otimes H^*$, corresponding to the Heisenberg equation. After obtaining the time evolved MPS, one can split it back to MPO and calculate $F(r, t)$.

\begin{figure}[t]
\centering
\includegraphics[width=0.5\textwidth=]{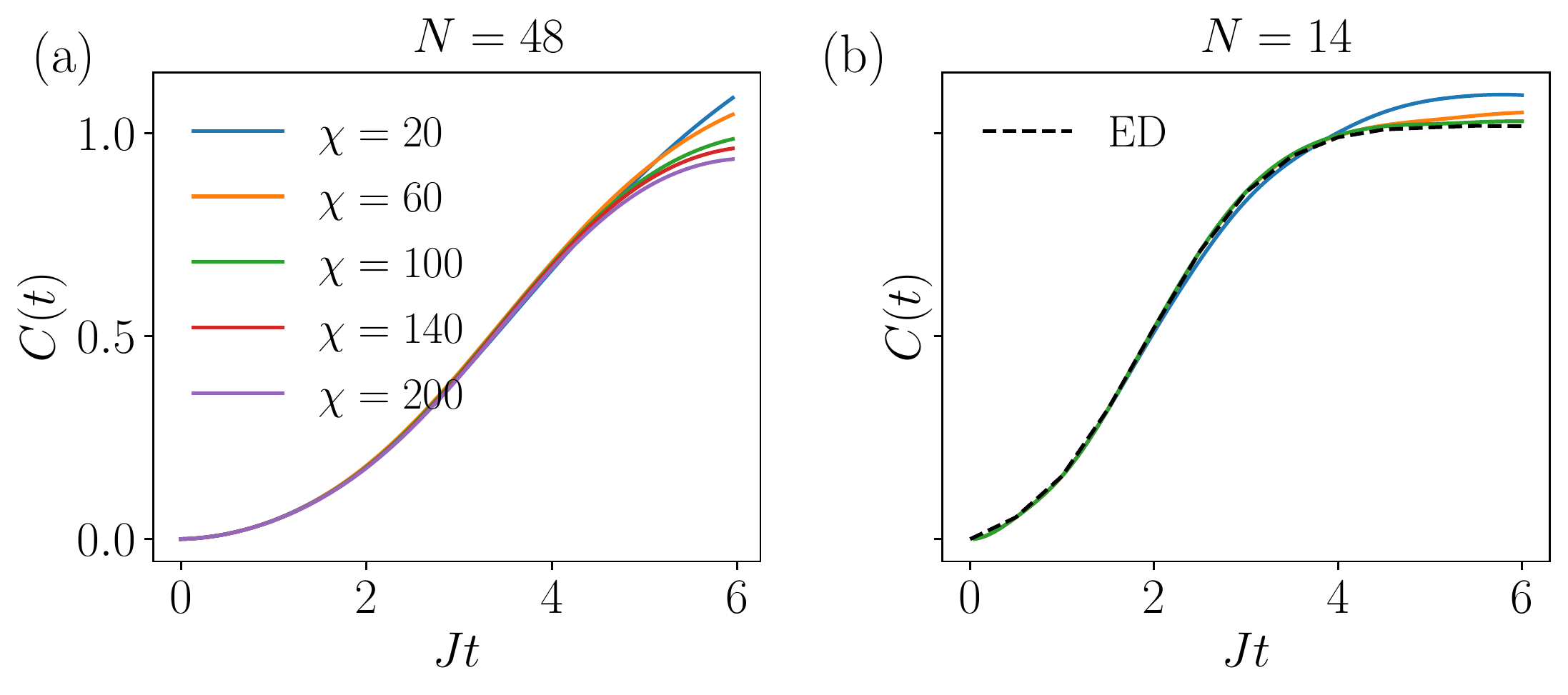}
\caption{The convergence of the numerical results in the tensor network simulation. (a) TDVP-MPO results with bond dimensions $\chi$ ranging from $20$ to $200$ for the largest system ($N=48$) studied in this work. (b) Comparison of the results obtained by MPO-TDVP method (solid) and ED (dashed) for a system with $N=14$.}
\label{fig:extfig3}
\end{figure}

In numerical simulations, we find that the bond dimension of the MPO $\chi=100$ is large enough to obtain the dynamics up to $N=48$ sites within the truncation error $10^{-5}$. We present the results obtained with $\chi$ ranging from 20 to 200 in Fig.~\ref{fig:extfig3} (a) for the largest system $N=48$ in our work, which show $C(t)$ converges quickly when increasing the bond dimension $\chi$. The result for $\chi = 100$ is accurate at least up to $Jt\sim 5.0$. In Fig.~\ref{fig:extfig3}(b) we compare the results obtained by TDVP-MPO and exact diagonalization (ED) for a small system with $N=14$. The TDVP-MPO data already converges to the ED results well with a small bond dimension $\chi$. 

The high accuracy and fast convergence in the TDVP-MPO method are attributed to the fact that it uses the Heisenberg picture and the accuracy is related to the operator entanglement, which is very different from the entanglement in pure states. It was shown the operator entanglement generated during the scrambling dynamics is quite limited and one can obtain accurate results of the OTOC for a large system (200 sites) with a small bond dimension~\cite{xu_accessing_2020}. It is also accurate for long-range interacting chaotic models~\cite{zhou_operator_2020}.

\section{Scrambling dynamics for $\alpha \neq 0.5$}\label{other_alpha}

\begin{figure}[t]
  \centering
  \includegraphics[width=0.35\textwidth=]{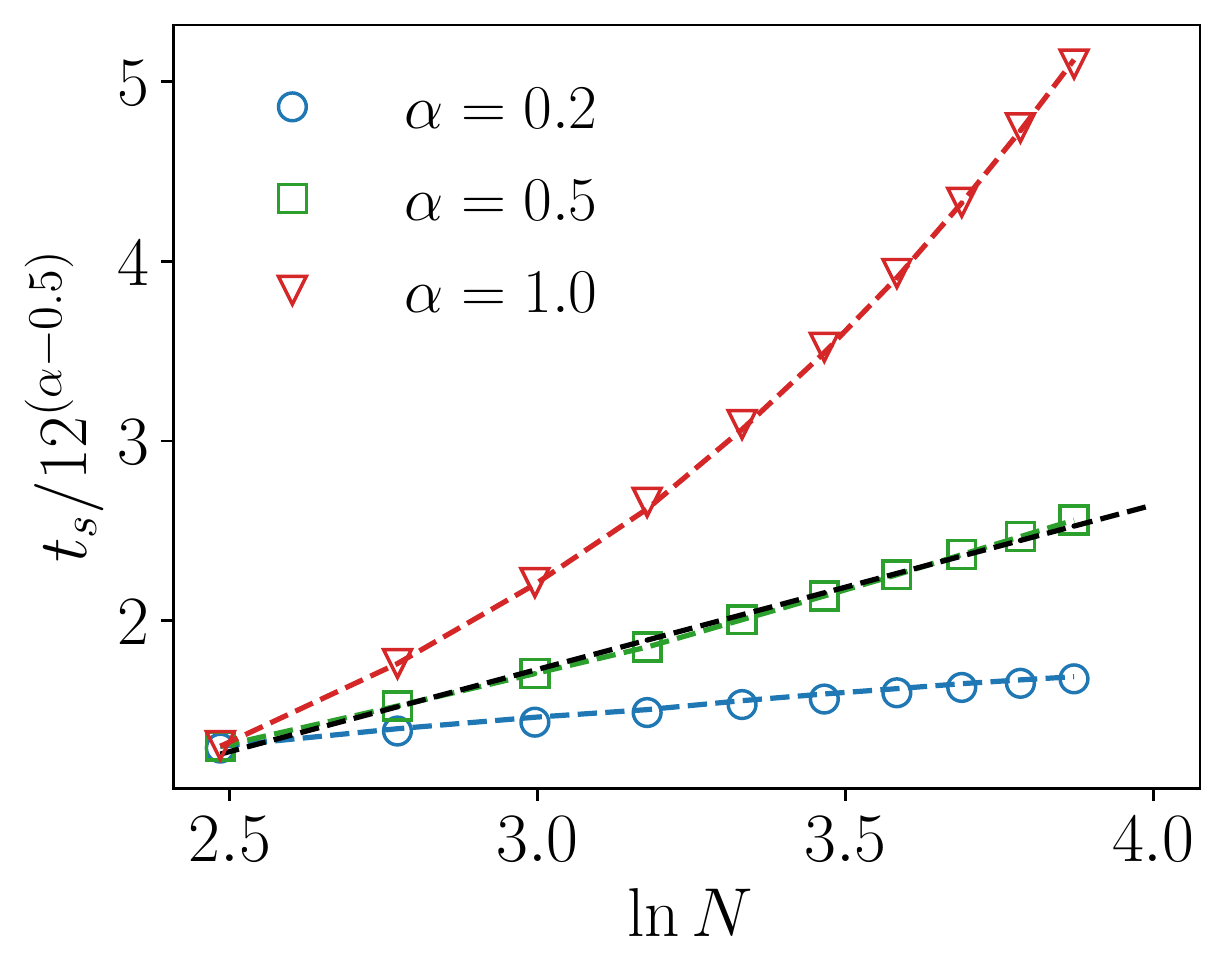}
  \caption{The fast scrambling for Hamiltonian (\ref{Hamil}) with different $\alpha$. Fast scrambling time for $\alpha=0.2, 0.5, 1.0$. The disorder strength are chosen as $W=4.0, 2.0, 0.55$ respectively such that the model are in the thermal phase. The discrete points are derived by direct simulation using Hamiltonian (\ref{Hamil}) while the dashed line are derived by rescaling the result for $\alpha=0.5$. Different $t_s$-$N$ curves are rescaled by a $N$ independent factor $12^{\alpha-0.5}$  for better presentation. }
  \label{fig:extfig2}
  \end{figure}

We have shown the fast scrambling dynamics for $\alpha=0.5$. In this section, we study how the scrambling dynamics vary with $\alpha \neq 0.5$. Before proceeding, we must note that with fixed disorder strength $W$, the system is unstable in any phase for $\alpha \neq 2$ as increasing $N$ (see Sec. ~\ref{transition}). Thus, Hamiltonian ~(\ref{Hamil}) may lead to different phases for small $N$ and large $N$ when studying the fast scrambling.
To study the dynamics in different non-equilibrium phases, it is neccessary to define the following Hamiltonian
\begin{equation}\label{Hamil2}
  \hat{H} = \frac{J}{2N^{\alpha}}\sum_{i<j} \left( \hat{\sigma}^{x}_i \hat{\sigma}^{x}_j + \hat{\sigma}^{y}_i \hat{\sigma}^{y}_j \right) +  \frac{1}{N^{\alpha-\tilde{\alpha}}} \sum_i h_i\hat{\sigma}^x_i
\end{equation}
Here, the $\tilde{\alpha}$ rescale the disorder strength relative to the interaction terms while the global interaction is tuned by $\alpha$. Note that the difference between Hamiltonian Eq.~(\ref{Hamil}) and Eq.~(\ref{Hamil2}) is just a global rescaling $1/N^{\alpha-\tilde{\alpha}}$. This leads to another fact that with fixed the disorder strength, i.e. $W$ and $\tilde{\alpha}$, the dynamics for different global interactions $\alpha_1, \alpha_2$ can be related by simply rescaling the Hamiltonian with a factor $N^{\alpha_1-\alpha_2}$. In this way, the scrambling time for different $\alpha$ can also be derived by rescaling.   
 However, it is interesting to ask whether the original model Eq.~(\ref{Hamil}) exhibits similar behavior, where the scrambling time for $\alpha \neq 0.5$ can not be deduced by rescaling the time. To this end, we obtain the scrambling time in two different ways. One is to directly change the $\alpha$ in the Hamiltonian (\ref{Hamil}), and the other is to calculate the scrambling time by rescaling the results of $\alpha=0.5$. The results are shown in Fig.~\ref{fig:extfig2}. We can see that for Hamiltonian (\ref{Hamil}) the dependence of fast scrambling behavior via $\alpha$ is similar to simply rescaling the time scale. Namely the scrambling times for $\alpha \neq 0.5$ are consistent with $t_s\sim N^{\alpha-0.5}\log N$, even without the directly rescaling equivalence property.  This may be because there are no other interaction terms in our model except the terms related to $\alpha$, which lead to the simple relation between the scrambling dynamics for different $\alpha$. Thus, our model shows a faster (slower) scrambling speed for $\alpha<0.5$ ($\alpha>0.5$), which seems to violate the fast scrambling conjecture. 

Actually, the faster and slower types of scrambling dynamics can exist in quantum lattice models. For instance, the quantum lattice models on a star graph~\cite{lucas_quantum_2019} can have a constant scrambling time, which can be seen as the fastest scrambling. On the other hand, for the slower scrambling region, our results are in agreement with the analytical result $t_s \gtrsim N^{\alpha-0.5}$~\cite{yin_bound_2020} that the fast scrambling is absent for $\alpha>0.5$. However, the results in Fig.~(\ref{fig:extfig2}) motivate us to introduce a proper timescale for the fast scrambling at infinite temperature.
At finite temperature, the inverse temperature $\beta$ serves as a natural timescale, the fast scrambler conjecture can be stated as the Lyapunov exponent is upper bounded by $2\pi$. In the infinite temperature calculation, however, there is no such a natural scale. In analogy to the finite temperature ensemble, we can introduce a proper timescale and one should study the information scrambling via the dimensionless $t/t_0$. For our model,  $t_0$ can naturally be chosen as $t_0 \sim N^{\alpha-0.5}$. It seems that using this time scale there exists fast scrambling for all $\alpha$. However, note that the timescale $t_0$ should not be divergent for large $N$. It is only valid for the region $\alpha < 0.5$ where $t_0 \rightarrow 0$ for large $N$ consisting with the infinite temperature $\beta=0$. Thus, we conclude that for $\alpha \leq0.5$ our model is a fast scrambler, in which the rescaled scrambling time $t_s/t_0 \sim \log N$.

\end{document}